\documentclass[twocolumn,english,showpacs,preprintnumbers,prb]{revtex4-1}
\usepackage{amssymb}
\usepackage[dvips]{color}
\usepackage{epsfig}
\usepackage{dcolumn}
\usepackage{bm}
\usepackage{graphicx,color}

\def\a{\alpha}

\def\G{\Gamma}
\def\d{\delta}

\def\e{\epsilon}
\def\ve{\varepsilon}

\def\t{\tau}

\def\w{\omega}

\def\l{\lambda}

\def\S{\Sigma}
\def\wid{\widetilde}

\begin{document}

\title{Nonequilibrium transport in a quantum dot attached to a Majorana bound state}

\author{S. J. S. da Silva,$^1$ A. C. Seridonio,$^{2,3}$ J. Del Nero,$^4$} \author{F. M. Souza$^5$}\email{Corresponding author: fmsouza@infis.ufu.br}
\affiliation{$^1$P\'os-Gradua\c{c}\~ao em F\'isica, Universidade Federal do Par\'a, 66075-110, Bel\'em, PA, Brazil\\
$^2$Instituto de Geoci\^{e}ncias e Ci\^{e}ncias Exatas, Universidade Estadual Paulista, Departamento de F\'{i}sica, 13506-970, Rio Claro,
SP, Brazil\\
$^3$Departamento de F\'{i}sica e Qu\'{i}mica, Universidade Estadual Paulista, 15385-000, Ilha Solteira, SP, Brazil\\
$^4$Departamento de F\'isica, Universidade Federal do Par\'a, 66075-110, Bel\'em, PA, Brazil\\
$^5$Instituto de F\'isica, Universidade Federal de Uberl\^andia, 38400-902, Uberl\^{a}ndia, MG, Brazil}

\begin{abstract}
We investigate theoretically nonequilibrium quantum transport in a
quantum dot attached to a Majorana bound state. Our approach is based
on the Keldysh Green's function formalism, which
allows us to investigate the electric current continuously
from the zero-bias limit up to the
large bias regime. In particular, our findings fully agree with previous results
in the literature that calculate transport using linear response theory
(zero-bias) or the master equation (high bias). Our
$I-V$ curves reveal a characteristic slope given by $I=(G_{0}/2)V$ in linear response regime,
where $G_0$ is the ballistic conductance $e^{2}/h$
as predicted in Phys. Rev. B 84, 201308(R) (2011).
Deviations from this behavior is also discussed when the dot couples
asymmetrically to both left and right leads. The differential conductance
obtained from the left or the right currents can be larger or smaller
than $G_{0}/2$ depending on the strength of the
coupling asymmetry. In particular, the standard conductance derived
from the Landauer-B\"{u}ttiker equation in linear response regime does
not agree with the full nonequilibrium calculation, when the two leads
couple asymmetrically to the quantum dot. We also compare the current through
the quantum dot coupled to a regular fermionic (RF) zero-mode or to
a Majorana bound state (MBS). The results differ considerably for the entire bias voltage
range analyzed. Additionally, we observe the formation of a plateau
in the characteristic $I-V$ curve for intermediate bias voltages when the
dot is coupled to a MBS. Thermal effects are also considered. We note
that when the temperature of the reservoirs is large enough both RF and MBS cases
coincide for all bias voltages.
\end{abstract}


%\volumeyear{year} \volumenumber{number} \issuenumber{number}
%\eid{identifier}
%\date[Date: ]{\today}

\maketitle

\section{Introduction}

In 1937 Ettore Majorana realized that the Dirac
equation could be modified to support a new class of particles called Majorana fermions
(MFs), with the intriguing property
that these particles are their own anti-particles.\cite{em37,JA12} Mathematically,
if $\eta$ is the annihilation operator for a Majorana particle then
$\eta=\eta^{\dagger}$.
These exotic particles are non-abelian anyons,
which means that particle exchanges are not merely accompanied by
a $+1$ for bosons or a $-1$ for fermions that multiplies the wave
function. Additionally, the exchange statistics of MFs
does not follow the regular anyons observed as quasiparticles in 2D
systems, where the exchange operation yields a Berry phase $e^{i\phi}$
multiplying the wave function.\cite{ml12}
Thus we end up for MFs
with an exotic non-Abelian exchange statistics. Until now no elementary
particle on nature was found as a MF. There is one possibility that
neutrinos might be MFs. On going experiments are attempting to verify this
hypothesis.\cite{fw09} Despite its origin in high energy physics, MFs came
recently in the news as a quasi-particle excitation in the low-energy
field of solid state physics.\cite{ja13}

Thus in the last few years the pursuit for devices
hosting MFs has received much attention from the scientific community,
in particular working with quantum computing. Such a quest is due
to the possibility of bounding two far apart MFs in order to define a nonlocal
qubit completely immune to the decoherence effect, which is crucial for the accomplishment of a robust topological quantum
computer.\cite{ML11,Lu12,MLKF12,KF12,MLKF2012} To this end, experimental
realizations should reveal first signatures of MFs that ensure the existence of them and hence,
their application as essential blocks for quantum computing. Nowadays, the most
promising setups for this goal lies on the superconductor based systems.\cite{AYK01,MG12,Lang12,Lin12,Liu13,DS13,JLiu12,SN13,DR12}

For instance, it was recently measured as a MF signature
a zero-bias peak in the conductance between a normal metal and the
end of a semiconductor nanowire (InSb) that is attached to a s-wave
superconductor.\cite{vm12,mtd12} This superconductor induces superconductivity
in the InSb nanowire via the proximity effect. In the presence of
a magnetic field parallel to the wire it was found a peak sticked
at the midgap of the nontrivial topological superconductor. This peak
is washed out for zero magnetic fields or when the magnetic field
is parallel to the spin-orbit field of the wire. Additionally, this
peak tends to disappear when temperature increases. All these features
are in agreement with theoretical works that settle the ingredients
necessary to have a Majorana bound state (MBS) in
a hybrid nanowire-superconductor device.\cite{DLoss,rml10,yo10,ktl09,jds10}
However, an alternative explanations for these measurements
were later proposed.\cite{tds12} Additionally,
MF are expected to appear in a variety of solid state systems, namely,
topological superconductors\cite{JA12} and fractional quantum Hall systems\cite{nr00}.
Moreover, MFs are theoretically predicted to appear in a half-quantum vortex of
a p-wave superconductors\cite{DA01} or at the ends of supercondutor vortices in doped topological insulators.\cite{PPRA11}

In solid state physics the main way to probe
MBSs is via conductance. A few experiments use tunneling
spectroscopy to probe MFs as a zero-bias anomaly.
There are some theoretical proposals that deal with transport through
a single level quantum dot attached to a left and to a right lead
and to a MBS in the end of a quantum wire.\cite{del11,EV13}
The main transport feature found for this system is a
conductance peak pinned at zero-bias with an amplitude of one-half
the ballistic conductance $G_{0}=e^{2}/h$,\cite{del11} valid when
the left and right leads couple symmetrically to the dot. We
point out that in Ref. {[}\onlinecite{EV13}{]}, E. Vernek \textit{et
al.} have found that such a value arises from the
leaking of the MBS into the quantum dot. Additionally,
the transport in this system was investigated in the large bias regime,
revealing a non-conserving current between left and
right leads.\cite{yc12}

In the present paper we apply the Keldysh nonequilibrium Green's function
technique\cite{hh08} to extend these previous works to the whole bias voltage
window, ranging from the zero-bias limit up to the large
bias regime. So, instead of focusing only on the zero-bias anomaly,
we explore the whole $I-V$ curve in the presence of a
single MBS. For comparison we also show the results to the case of
a regular fermionic (RF) zero-mode coupled to the dot. Both cases (MBS and RF)
differ appreciably along the bias voltage window, not only in the
zero-bias regime. We observe, for instance,
the formation of an additional plateau in the I-V curve when the dot is coupled
to a MBS. Additionally, it is found
a slope at the characteristic $I-V$ curve equal to one-half the quantum of conductance 
$G_{0}$ when the bias voltage tends to zero, in accordance to Ref.
{[}\onlinecite{del11}{]}.

We pay particular attention to the coupling asymmetry between \emph{left lead-quantum dot} and \emph{right
lead-quantum dot}. These couplings are characterized by the tunneling rates $\G^L$ and $\G^R$, respectively.
We investigate the cases $\G^L > \G^R$ and $\G^L < \G^R$. Cao et al.\cite{yc12} found that in the
large bias regime the current is not conserved with $I_L > I_R$ or $I_L < I_R$ depending on the
asymmetry factor $y=\G^R/\G^L$. Interestingly, the nonconserving feature also affects the zero-bias
conductance. The zero-bias limit departs from $G_{0}/2$ when the leads couple asymmetrically ($y \neq 1$) to the dot.
We have found in the zero-bias limit $dI_{L}/dV>G_{0}/2$ and $dI_{R}/dV<G_{0}/2$
or the opposite, depending on the degree of asymmetry $y$. Neither $dI_{L}/dV$
nor $dI_{R}/dV$ coincide with the conductance obtained via the Landauer-B\"{u}ttiker
equation in linear response regime, except for symmetric couplings ($y = 1$).
This indicates that a full nonequilibrium quantum transport formulation
is more suitable to describe the system with Majorana bound state.
Thermal effects are also investigated. We observe that when the temperature is large enough both
MBS and RF cases become indistinguishable for any bias voltage.

The paper is organized as follows. In Sec. II we present a detailed
derivation of the nonlinear transport equations obtained via Keldysh
technique. In Sec. III we show the main results found and in Sec.
IV we conclude.

\begin{figure}[!htb]
\begin{center}
\includegraphics[height=6cm]{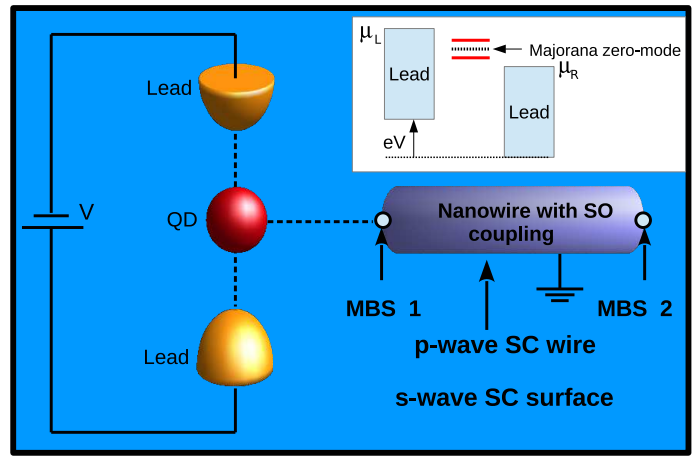}
\caption{(Color online) Sketch of the system studied. A single level quantum dot is coupled to both
left and right leads. In the presence of a bias voltage the system is driven
away the equilibrium and a tunnel current passes through
the dot. A semiconductor quantum wire (e.g. InSb) with strong spin orbit
interaction lies on an s-wave superconductor that induces a topological phase 
in the wire, resulting in localized Majorana states
at the ends of this wire, which is crossed by an applied 
magnetic field parallel to it and perpendicular to its spin-orbit field. The Majorana bound
state closer to the dot couples to it, which can affect the characteristic $I-V$ profile.}\label{fig1}
\end{center}
\end{figure}

\section{Model and Formulation}
To describe the system presented in Fig. (\ref{fig1}) we use the
Hamiltonian originally proposed by Liu and Baranger,\cite{del11}
\begin{eqnarray}
 H &=& H_{\text{leads}}+H_{\text{dot}}+H_{\text{T}}+ \e_M (f^\dagger f - \frac{1}{2}) + \nonumber \\ &&
 \frac{\lambda_A}{\sqrt{2}} (d f^\dagger + f d^\dagger) +\frac{\lambda_B}{\sqrt{2}} (d f + f^\dagger d^\dagger),
\end{eqnarray}
where the first term gives the free-electron energy of the reservoirs,
$H_{\text{leads}}=\sum_{k,\alpha}\varepsilon_{k}c_{k,\alpha}^{\dag}c_{k,\alpha}$,
the second term is the single level quantum dot Hamiltonian, $H_{dot}=\varepsilon_{d}d^{\dag}d$
and the third term gives the tunnel coupling between the quantum dot
and the leads, $H_{\text{T}}=\sum_{k,\alpha}[V_{\alpha}c_{k,\alpha}^{\dag}d+V_{\alpha}^{*}d^{\dag}c_{k,\alpha}]$,
with $\a=L$ (left lead) or $\a=R$ (right lead).
The fourth term accounts for the Majorana modes, and the last two
terms can be understood as follows: (i) $\lambda_{A} = \lambda \neq 0$, $\lambda_{B}=0$ and $\e_{M}=0$
we have a regular fermionic (RF) zero-mode attached to the quantum dot and (ii) for
$\lambda_{A}=\lambda_{B}=\lambda$ we obtain a MBS coupled to the
quantum dot. In case (i) the Hamiltonian becomes
\begin{equation}
 H = H_{\text{leads}}+H_{\text{dot}}+H_{\text{T}}+\frac{\lambda}{\sqrt{2}} (d f^\dagger + f d^\dagger),
\end{equation}
while for (ii) we have 
\begin{equation}
 H=H_{\text{leads}}+H_{\text{dot}}+H_{\text{T}}+i \e_M \eta_1 \eta_2 +\lambda (d-d^\dagger) \eta_1,
\end{equation}
where $\eta_1=\frac{f^\dagger+f}{\sqrt{2}}$ and $\eta_2=i\frac{f^\dagger-f}{\sqrt{2}}$.
In the following nonequilibrium calculation we consider this last Hamiltonian. In order to compare our findings with 
the ones obtained previously for the large bias limit, we adopt $\lambda=\sqrt{2} \lambda'$, where 
$\lambda'$ gives the tunnel coupling between the dot and the nearby MBS in Ref. [\onlinecite{yc12}].
We highlight that the present spinless Hamiltonian for MFs assumes
a strong magnetic field applied on the whole setup of Fig. (\ref{fig1}), thus resulting a large
Zeeman splitting where the higher levels are not energetic favorable
within the operational temperatures of the system. In this case, one spin component becomes completely inert
and the spin degrees of freedom can be safely ignored. As a result, the Coulomb interaction between opposite
spins in the quantum dot is avoided and the model becomes exactly solvable.
To our best knowledge, this work is the first to obtain such a solution by using Green's functions in the Keldysh framework.

The current in the lead $\alpha$ can be calculated
from the definition $I_{\a}=-e\langle\dot{N}_{\a}\rangle$, where
$e>0$ is the modulus of the electron charge. $N_{\a}=\sum_{k}c_{k\a}^{\dagger}c_{k\a}$
is the total number operator for lead $\a$ and $\langle...\rangle$
is a thermodynamics average. The time derivative of $N_{\a}$ is calculated
via Heisenberg equation, $\dot{N}_{\a}=i[H,N_{\a}]$ (we adopt $\hbar = 1$), which results
in\cite{hh08}
\begin{equation}
I_{\a}=2e\mathrm{Re}[\sum_{k_{a}}V_{\a}G_{d,k_{\a}}^{<}(t,t)],
\end{equation}
where $G_{d,k_{\a}}^{<}(t,t)=i\langle c_{k_{\a}}^{\dagger}(t)d(t)\rangle$.
After a straightforward calculation the current expression can be
cast into the following form
\begin{equation}
I_{\a}=ie\int\frac{d\w}{2\pi}\G_{\a}\{[G_{d}^{r}(\w)-G_{d}^{a}(\w)]f_{\a}+G_{d}^{<}(\w)\}.\label{I}
\end{equation}
Here $\G_{\a}=2\pi V_{\alpha}^2\rho_{\a}$, with $\rho_{\a}$ being the density
of states of the reservoir $\a$, and the Green's functions
$G_{d}^{r}(\w)$, $G_{d}^{a}(\w)$ and $G_{d}^{<}(\w)$ are the retarded,
advanced and lesser Green's functions of the quantum
dot. These Green's functions can be obtained via analytic continuation
of the contour-ordered Green's functions $G_{d}(\t,\t')=-i\langle T_{c}d(\t)d^{\dagger}(\t')\rangle$,
where $T_{c}$ orders the operators along the Keldysh contour. Since
the equation of motion for $G_{d}(\t,\t')$ is structurally equivalent
to the chronological time-ordered Green's function
$G_{d}(t,t')=-i\langle T d(t)d^{\dagger}(t')\rangle$,\cite{hh08} in what follows
we calculate $G_{d}(t,t')$ via equation of motion technique. Taking
the time derivative with respect to $t$ we obtain
\begin{eqnarray}
[i\frac{\partial}{\partial t}-\varepsilon_{d}]G_{d}(t,t') & = & \delta(t-t')+\sum_{k,\alpha}V_{\alpha}^{*}G_{c_{k\a}}(t,t')\nonumber \\
 &  & \phantom{xxxxxx}-\lambda G_{\eta_{1}}(t,t'),\label{Gd}
\end{eqnarray}
where the additional Green's functions were defined
as $G_{c_{k\a}}(t,t')=-i\langle Tc_{k\alpha}(t)d^{\dagger}(t')\rangle$
and $G_{\eta_{1}}(t,t')=-i\langle T\eta_{1}(t)d^{\dagger}(t')\rangle$.
Calculating the time-derivative of these new Green's
function with respect to $t$ we find
\begin{eqnarray}
[i\frac{\partial}{\partial t}-\varepsilon_{k,\a}]G_{c_{k\a}}(t,t')=V_{\a}G_{d}(t,t'),\label{Gk}
\end{eqnarray}
and
\begin{eqnarray}
i\frac{\partial}{\partial t}G_{\eta_{1}}(t,t')=i\varepsilon_{M}G_{\eta_{2}}(t,t')-\lambda G_{d}(t,t')+\lambda G_{d^{\dagger}}(t,t').\nonumber \\
\label{Geta1}
\end{eqnarray}
Observe that two new Green's functions arise at this
last equation, namely, $G_{\eta_{2}}(t,t')=-i\langle T\eta_{2}(t)d^{\dagger}(t')\rangle$
and $G_{d^{\dagger}}(t,t')=-i\langle Td^{\dagger}(t)d^{\dagger}(t')\rangle$.
Performing once again the time-derivative with respect to $t$ of
these two Green's functions we arrive at
\begin{eqnarray}
i\frac{\partial}{\partial t}G_{\eta_{2}}(t,t')=-i\varepsilon_{M}G_{\eta_{1}}(t,t'),\label{Geta2}
\end{eqnarray}
and
\begin{eqnarray}
[i\frac{\partial}{\partial t}+\varepsilon_{d}]G_{d^{\dag}}(t,t')=-\sum_{k,\alpha}V_{\alpha}G_{c_{k\alpha}^{\dagger}}(t,t')+\lambda G_{\eta_{1}}(t,t').\nonumber \\
\label{Gddagger}
\end{eqnarray}
One more Green's function appears at this last results,
$G_{c_{k\alpha}^{\dagger}}(t,t')=-i\langle Tc_{k\alpha}^{\dagger}(t)d^{\dagger}(t')\rangle$,
whose equation of motion can be easily calculated,
\begin{eqnarray}
[i\frac{\partial}{\partial t}+\varepsilon_{k\alpha}]G_{c_{k\alpha}^{\dagger}}(t,t')=-V_{\alpha}^{*}G_{d^{\dag}}(t,t').\label{Gkdagger}
\end{eqnarray}
Equations (\ref{Gd}), (\ref{Gk}), (\ref{Geta1}), (\ref{Geta2}),
(\ref{Gddagger}) and (\ref{Gkdagger}) constitute
a complete set of six differential equations. In order to reduce to
only four equations we write Eqs. (\ref{Gk}) and (\ref{Gkdagger})
in their integral forms\cite{comment1}
\begin{eqnarray}
G_{c_{k\a}}(t,t') & = & V_{\a}\int dt_{1}g_{k\a}(t,t_{1})G_{d}(t_{1},t'),\\
G_{c_{k\alpha}^{\dagger}}(t,t') & = & -V_{\alpha}^{*}\int dt_{1}g'_{k\a}(t,t_{1})G_{d^{\dag}}(t,t')\label{Gckadagger}
\end{eqnarray}
and use them into Eqs. (\ref{Gd}) and (\ref{Gddagger}). This gives
us
\begin{eqnarray}
[i\frac{\partial}{\partial t}-\varepsilon_{d}]G_{d}(t,t') & = & \delta(t-t')+\int dt_{1}\S(t,t_{1})G_{d}(t_{1},t')\nonumber \\
 &  & \phantom{xxxxxx}-\lambda G_{\eta_{1}}(t,t'),\label{Gdint}
\end{eqnarray}
and
\begin{eqnarray}
[i\frac{\partial}{\partial t}+\varepsilon_{d}]G_{d^{\dag}}(t,t') & = & \int dt_{1}\S'(t,t_{1})G_{d^{\dag}}(t_{1},t')\nonumber \\
 &  & \phantom{xxxxxx}+\lambda G_{\eta_{1}}(t,t'),\nonumber \\
\label{Gddaggerint}
\end{eqnarray}
where $\S(t,t_{1})=\sum_{k\a}|V_{\a}|^{2}g_{k\a}(t,t_{1})$ and $\S'(t,t_{1})=\sum_{k\a}|V_{\a}|^{2}g'_{k\a}(t,t_{1})$.
Equations (\ref{Geta1}), (\ref{Geta2}), (\ref{Gdint}) and (\ref{Gddaggerint})
constitute our new set of four-integrodifferential equations, which
can be written in a matrix form as
\begin{widetext}
\begin{eqnarray}
 \left[
\begin{array}{cccc}
    i \frac{\partial}{\partial t}-\ve_d & 0 & 0 & 0 \\
    0& i \frac{\partial}{\partial t} & 0 & 0 \\
    0 & 0 & i \frac{\partial}{\partial t} & 0 \\
    0 & 0 & o & i \frac{\partial}{\partial t}+\varepsilon_{d} \\
    \end{array} \right]
    \left[
\begin{array}{c}
    G_{d}(t,t')  \\
    G_{\eta_{1}}(t,t')   \\
    G_{\eta_{2}}(t,t')  \\
    G_{d^{\dagger}}(t,t') \\
    \end{array} \right] &=&\delta(t-t')\left[
\begin{array}{c}
    1 \\
    0 \\
    0 \\
    0 \\
    \end{array} \right]
    +\int dt_{1}\left[
\begin{array}{cccc}
    \S(t,t_{1}) & 0 & 0 & 0 \\
    0& 0 & 0 & 0 \\
    0 & 0 & 0 & 0 \\
    0 & 0 & 0 &\S'(t,t_{1})  \\
    \end{array} \right]
    \left[
\begin{array}{c}
    G_{d}(t_1,t')   \\
    G_{\eta_{1}}(t_1,t')    \\
    G_{\eta_{2}}(t_1,t')   \\
    G_{d^{\dagger}}(t_1,t')  \\
    \end{array} \right]+\nonumber\\&&
    \phantom{xxxxxxxxxxxx}
    \left[
\begin{array}{cccc}
    0 & -\lambda & 0 & 0 \\
   -\lambda& 0 & i\varepsilon_{M} & \lambda \\
    0 & -i\varepsilon_{M} & 0 & 0 \\
    0 & \lambda & 0 & 0 \\
\end{array} \right]
\left[
\begin{array}{c}
    G_{d}(t,t')  \\
    G_{\eta_{1}}(t,t')   \\
    G_{\eta_{2}}(t,t')  \\
    G_{d^{\dagger}}(t,t') \\
    \end{array} \right],
\end{eqnarray}
\end{widetext}
or in a more compact way as
\begin{eqnarray}\label{Gvec}
 \vec{G}(t,t')=\mathbf{g}(t,t') \vec{u} + \int \int dt_1 dt_2 \mathbf{g}(t,t_1) \wid{\mathbf{\S}}(t_1,t_2)\vec{G}(t_2,t'),\nonumber\\
\end{eqnarray}
where the matrix $\mathbf{g}(t,t')$ is defined according to
\begin{eqnarray}
 \left[
\begin{array}{cccc}
    i \frac{\partial}{\partial t}-\ve_d & 0 & 0 & 0 \\
    0& i \frac{\partial}{\partial t} & 0 & 0 \\
    0 & 0 & i \frac{\partial}{\partial t} & 0 \\
    0 & 0 & o & i \frac{\partial}{\partial t}+\varepsilon_{d} \\
    \end{array} \right]  \mathbf{g}(t,t')=\d(t-t') \mathbf{I},\nonumber\\
\end{eqnarray}
with $\mathbf{I}$ being the $4 \times 4$ identity matrix, and
\begin{eqnarray}
 \wid{\mathbf{\S}}(t,t')&=&\left[
\begin{array}{cccc}
    \S(t,t') & 0 & 0 & 0 \\
    0& 0 & 0 & 0 \\
    0 & 0 & 0 & 0 \\
    0 & 0 & 0 &\S'(t,t')  \\
    \end{array} \right]+\nonumber\\ && \d(t-t')\left[
\begin{array}{cccc}
    0 & -\lambda & 0 & 0 \\
   -\lambda& 0 & i\varepsilon_{M} & \lambda \\
    0 & -i\varepsilon_{M} & 0 & 0 \\
    0 & \lambda & 0 & 0 \\
\end{array} \right].
\end{eqnarray}
The vectors $\vec{G}$ and $\vec{u}$ are defined as
\begin{equation}
 \vec{G}(t,t')=\left[
\begin{array}{c}
    G_{d}(t,t')  \\
    G_{\eta_{1}}(t,t')   \\
    G_{\eta_{2}}(t,t')  \\
    G_{d^{\dagger}}(t,t') \\
    \end{array} \right]\phantom{xxx}\mathrm{and}\phantom{xxx}\vec{u}=\left[
\begin{array}{c}
    1 \\
    0 \\
    0 \\
    0 \\
    \end{array} \right].
\end{equation}
Iterating Eq. (\ref{Gvec}) we can show that
\begin{eqnarray}
  \vec{G}(t,t')=\mathbf{G}(t,t')\vec{u},
\end{eqnarray}
with the Dyson equation
\begin{eqnarray}
 \mathbf{G}(t,t')=\mathbf{g}(t,t') + \int \int dt_1 dt_2 \mathbf{g}(t,t_1) \wid{\mathbf{\S}}(t_1,t_2) \mathbf{G}(t_2,t').\nonumber\\
\end{eqnarray}
Writing a similar equation in the Keldysh contour,\cite{hh08}
\begin{eqnarray}\label{Gcontour}
 \mathbf{G}(\t,\t')=\mathbf{g}(\t,\t') + \int_C \int_C d\t_1 d\t_2 \mathbf{g}(\t,\t_1) \wid{\mathbf{\S}}(\t_1,\t_2) \mathbf{G}(\t_2,\t'),\nonumber\\
\end{eqnarray}
and applying the Langreth's analytical continuation rules,\cite{hh08} we obtain in the frequency domain
\begin{eqnarray}\label{GR}
 \mathbf{G}^r(\w)=\mathbf{g}^r(\w) + \mathbf{g}^r(\w) \wid{\mathbf{\S}}^r(\w) \mathbf{G}^r(\w),
\end{eqnarray}
to the retarded Green's function and
\begin{eqnarray}\label{G<}
 \mathbf{G}^<(\w)=\mathbf{G}^r(\w) {\mathbf{\S}}^<(\w) \mathbf{G}^a(\w),
\end{eqnarray}
to the lesser Green's function both already in the Fourier domain. The retarded and lesser components
of the self-energy can be expressed as
\begin{equation}
 \wid{\mathbf{\S}}^r(\w)=\left[
\begin{array}{cccc}
    -\frac{i}{2}\G(\w) & -\l & 0 & 0 \\
    -\l& 0 & i\ve_M & \l \\
    0 & -i\ve_M & 0 & 0 \\
    0 & \l & 0 & -\frac{i}{2}\G(-\w)  \\
    \end{array} \right],
\end{equation}
and ${\mathbf{\S}}^<(\w)$ has only two nonzero elements,
\begin{eqnarray}
 {{\S}}^<_{11}(\w) &=& i[\G_L(\w)f_L(\w)+\G_R(\w)f_R(\w)],\\
 {{\S}}^<_{44}(\w) &=& i[\G_L(-\w)f_L(-\w)+\G_R(-\w)f_R(-\w)],
\end{eqnarray}
With Eqs. (\ref{GR}) and (\ref{G<}) we can calculate the transport properties described below.

\begin{figure}[!htb]
\begin{center}
\includegraphics[height=5cm]{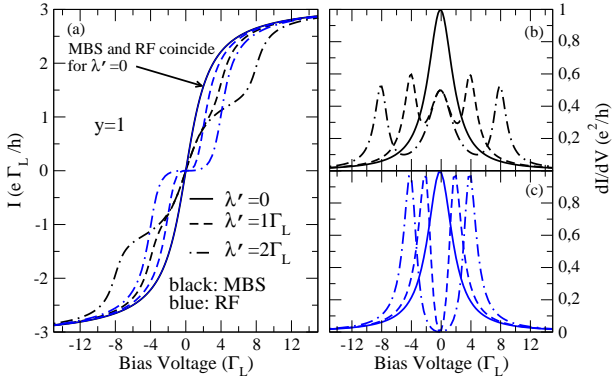}
\caption{(Color online) (a) Current and differential conductance
against the bias voltage in
units of $\G_L$ for differing $\lambda'$
and symmetric case $y=1$. Both MBS and RF cases
are shown, as black and blue lines, respectively. The $\lambda'=0$
gives the same results for both cases, which corresponds to a transport
through a single level quantum dot. For finite $\lambda'$ the two
cases present distinct $I-V$ profiles. The RF case shows a flat $I-V$
characteristics around zero bias and then it increases when the double-dot
conduction channels cross the reservoir chemical potential. Contrasting, the MBS regime yields a typical slope around zero
bias which turns into the $G_{0}/2$ as predicted in the literature.
In panels (b)-(c) we show $dI/dV$ for both cases. While in
the MBS the conductance is pinned at 0.5 $(\lambda'\neq 0)$ it is
zero in the RF regime. Parameters: $y=1$, $\G_R=y \G_L$, $k_BT=0.01\G_L$, 
$\e_M=0$, $\e_d=0$, $\lambda'=0$, $1\G_L$ and $2\G_L$, $\lambda=\sqrt{2}\lambda'$.}\label{fig2}
\end{center}
\end{figure}

\section{Results}

In Fig. \ref{fig2}(a) we compare the characteristic $I-V$ curve
in the cases of a MBS and a RF attached to the quantum dot.
We adopt $\G_L$ as our energy scale, so the bias voltage, the energy levels, and the coupling $\lambda'$
will be expressed in units of $\G_L$, while the currents in units of $e \G_L/ h$, with $h$ being the Planck's constant.
For $\lambda=0$ both results coincide and the system behaves
as a single level quantum dot. For $\lambda \neq 0$ distinct features
arise in each case. In particular, in the linear response regime,
the current presents a finite slope as the bias increases for the
MBS case while it is flat for the RF
situation. As the bias voltage increases above the linear response
regime, we observe the formation of a plateau in the current for the MBS case
and then it increases further, saturating at large enough
bias voltages. In contrast, for $\lambda_{B}=0$ (RF) we have a single step current profile,
without the formation of an intermediate plateau. For larger biases
the current coincides for both cases (RF and MBS).

\begin{figure}[!htb]
\begin{center}
\includegraphics[height=8cm]{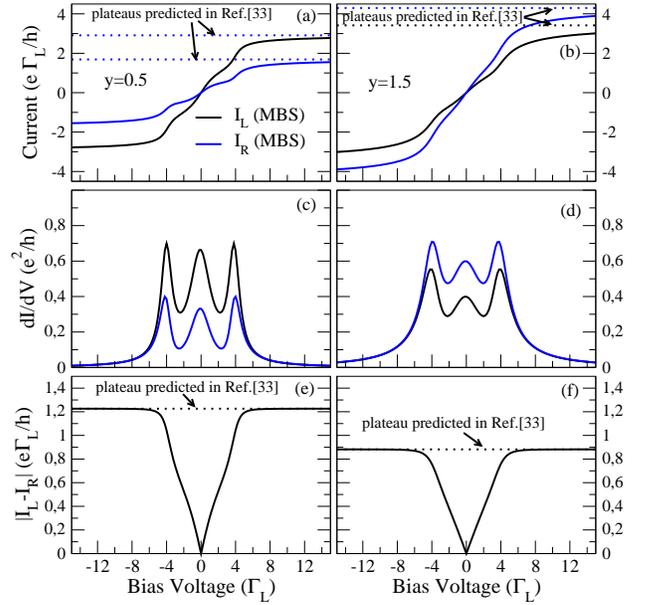}
\caption{(Color online) (a)-(b) Left and right lead currents,
(c)-(d) differential conductance $dI/dV$ and (e)-(f) current difference $|I_{L}-I_{R}|$
against the bias voltage in energy units of $\G_L$.
We consider $y=0.5$ (left panels) and $y=1.5$ (right panels). Both $I_{L}$ and $I_{R}$
present similar features against bias voltage but distinct values.
In particular the slope around zero bias and the plateaus differ from each other. In order to confirm our nonequilibrium
calculation we compare the high bias plateau with the ones predicted
by Cao \textit{et al.\cite{yc12}} via the
master equation technique. The different $|I_{L}-I_{R}|$ increases
with bias and then it saturates at the value predicted in the aforementioned
reference. The zero bias value of the differential conductance $dI/dV$
contrasts to the one obtained for the symmetric case $y=1$. Here
$dI_{L}/dV$ and $dI_{R}/dV$ are not at 0.5 and they differ from
each other, with $dI_{L}/dV>dI_{R}/dV$ for $y<1$ and the opposite
for $y>1$. Parameters: $\G_R=y \G_L$, $k_BT=0.01\G_L$, $\e_M=0$, $\e_d=0$, $\lambda'=1\G_L$.}\label{fig3}
\end{center}
\end{figure}

In Fig. \ref{fig2}(b) we show the differential conductance ($dI/dV$)
for the currents presented in Fig. \ref{fig2}(a) in the presence
of a MBS. For $\lambda=0$ the conductance $dI/dV$
is the standard Lorentzian with broadening given by $\G=\G_L+\G_R$. In contrast,
for $\lambda\neq0$ the conductance reveals a three peaks structure,
in which one of them has an amplitude of 0.5 pinned at zero-bias,
in accordance to the work of Liu and Baranger.\cite{del11}
For the RF, though, we find $dI/dV$ similar to the
characteristic T-shaped quantum dot geometry,\cite{ACS09} where the conductance is zero for bias voltage close
to zero. 

It is valid to note that the currents presented in Fig. (\ref{fig2})
for both MBS and RF cases can also be obtained from the standard Landauer-B\"{u}ttiker expression\cite{comment2}
\begin{equation}
I=\frac{e}{\hbar}\int\frac{d\w}{2\pi}[f_{L}(\w)-f_{R}(\w)]T(\w),\label{Isymmetric}
\end{equation}
where $T(\w)=[\G_{L}\G_{R}/(\G_{L}+\G_{R})](-2)\text{{Im}}[G_{dd}^{r}(\w)]$,
which gives the following conductance in the linear response limit
\begin{equation}
G=\frac{e^{2}}{h}\int d\w\frac{\G_{L}\G_{R}}{\G_{L}+\G_{R}}(-2)\text{{Im}}[G_{dd}^{r}(\w)][-\frac{\partial f}{\partial\w}].\label{Glinear}
\end{equation}
This symmetric expression is only true for charge conserving systems
where $I_{L}=-I_{R}$. This is always the case when $\lambda_{B}=0$
(RF). However, for $\lambda_{B}=\lambda_{A} \neq 0$ (MBS)
this is valid in the symmetric coupling regime ($y=1$) only. When
$y\neq1$ the left and right currents depart from each other, and
consequently the result obtained from Eq. (\ref{Isymmetric}) differs
from both $I_{L}$ and $I_{R}$ obtained via Eq. (\ref{I}).

In order to explore the coupling asymmetries ($y\neq 1$) in the transport, we
plot separately in Fig. (\ref{fig3}) both $I_{L}$ and $I_{R}$ for
$\lambda_{B}=\lambda_{A}=\lambda$ (MBS), and their corresponding $dI/dV$
profiles for two asymmetry factors $y=0.5$ (left panels) and $y=1.5$ (right panels). It is clear
from the plot that the system does not conserve current ($I_{L}\neq I_{R}$).
For larger enough bias voltages the currents $I_{L}$
and $I_{R}$ attain different plateaus, which are confirmed by the
analytical results, recently derived by Cao \emph{et al.} via Born-Markov
master equation technique, namely,\cite{yc12}
\begin{eqnarray}
I_{L} & = & \frac{\G_{L}\G_{R}}{\G}[1+\frac{4(1/y-1)\l'^{2}}{\G^{2}+4(\e_{d}^{2}+\e_{M}^{2}+2\l'^{2})}],\label{cao1}\\
I_{R} & = & \frac{\G_{L}\G_{R}}{\G}[1+\frac{4(y-1)\l'^{2}}{\G^{2}+4(\e_{d}^{2}+\e_{M}^{2}+2\l'^{2})}].\label{cao2}
\end{eqnarray}
These large bias limiting values are plotted in Fig. (\ref{fig3})
as dotted lines. Looking at the zero-bias limit,
one may note that the slopes of
$I_{L}$ and $I_{R}$ vs. $V$ deviate from each other with $|I_{L}|>|I_{R}|$
for $y=0.5$ and $|I_{L}|<|I_{R}|$ for $y=1.5$. The differential conductance
$dI_{L}/dV$ and $dI_{R}/dV$ clearly show the difference of the slopes
at zero bias, with $dI_{L}/dV\approx0.7$ and $dI_{R}/dV\approx0.3$
for $y=0.5$ and $dI_{L}/dV\approx0.4$ and $dI_{R}/dV\approx0.6$
for $y=1.5$. This contrasts with the symmetric case, where both conductances
are at 0.5, as predicted by Liu and Baranger.\cite{del11}

\begin{figure}[!htb]
\begin{center}
\includegraphics[height=5cm]{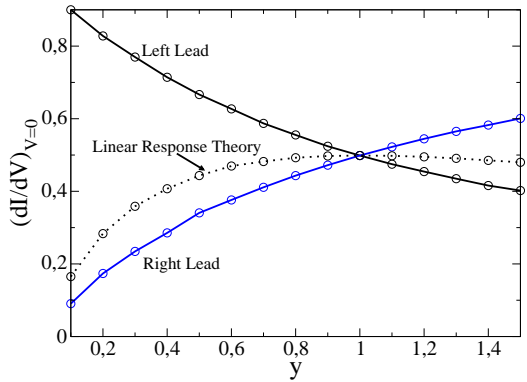}
\caption{(Color online) (a) Differential conductances
$dI_{L}/dV$ and $dI_{R}/dV$ at zero-bias against $y$ in the presence of a MBS.
$dI_{L}/dV$ is larger than $dI_{R}/dV$ for small $y$, they 
attain the same value at $y=1$ and then $dI_{R}/dV$ turns greater
than $dI_{L}/dV$ as $y$ becomes higher than one.
For comparison we show $dI/dV$ obtained via linear response theory. 
Parameters: $\G_R=y \G_L$, $k_BT=0.01\G_L$, $\e_M=0$, $\e_d=0$, $\lambda'=1\G_L$.}\label{fig4}
\end{center}
\end{figure}

\begin{figure}[!htb]
\begin{center}
\includegraphics[height=5cm]{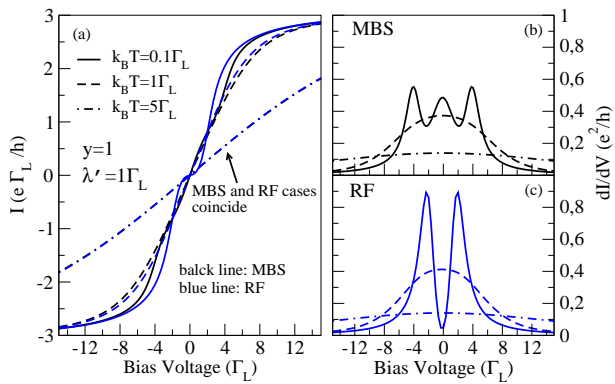}
\caption{(Color online) (a) Current and (b)-(c) differential
conductance against the bias voltage for three values of temperature: $k_B T=0.1\G_L$, 
$1\G_L$ and $5\G_L$. Both MBS and RF cases are shown. For small temperatures both cases
differ, however as $k_{B}T$ increases the two regimes tend to the same results. In particular,
the characteristic signature $dI/dV=0.5 G_0$ for a MBS is washed out
as the temperature enhances. Parameters: $y=1$, $\G_R=y \G_L$, $\e_M=0$, $\e_d=0$, $\lambda'=1\G_L$.}\label{fig5}
\end{center}
\end{figure}

In Fig. 3(e)-(f) we plot the difference $|I_{L}-I_{R}|$ against bias
voltage. It is clear that in the nonequilibrium regime the current
is not conserved with $I_{L}>I_{R}$ for $y<1$ and the opposite for
$y>1$. As the bias voltage enlarges and all the conduction channels
(three channels in the presence of a MBS) become
inside the conduction window, the difference $|I_{L}-I_{R}|$ attains
the plateau predicted by Eqs. (\ref{cao1})-(\ref{cao2}).

In Fig. (\ref{fig4}) we show how $dI/dV$ evolves with $y$ at the
zero-bias limit. Both $dI_{L}/dV$ (black) and $dI_{R}/dV$
(blue) are shown. As a matter of comparison we also plot $dI/dV$ obtained
via the standard linear response expression, Eq. (\ref{Glinear}).
While all results coincide for the symmetric case ($y=1$), they all
differ for $y\neq1$.

Finally, Fig. (\ref{fig5}) shows $I$ vs. $V$ curves and the corresponding
differential conductance for different temperatures in the symmetric
case ($y=1$). Both the MBS and RF cases are presented. As the temperature
increases the curves for both regimes tend to become smoother, as
expected due to the smearing out of the Fermi function
around the chemical potential of the electronic reservoirs. In particular,
opposite behavior between MBS and RF are seen at the slope of the $I-V$
curve around zero bias. While in the MBS the slope is suppressed for
increasing $k_{B}T$, it is amplified in the RF case for $k_B T=1 \G_L$. This behavior
can be clearly seen in the differential conductance $dI/dV$ at zero
bias. Remarkably, both MBS and RF cases coincide
for large enough temperature and the $I-V$ presents a linear profile.

\section{Conclusion}

We have studied nonequilibrium quantum transport
in a quantum dot attached to two leads and to a localized Majorana
bound state. Our approach, based on the Keldysh nonequilibrium Green's
function, allows us to study transport through the
whole bias voltage range, starting at the zero-bias limit and moving
up to the large bias regime. Previous works investigate separately
only the zero-bias or the large bias limit. To the best of our knowledge
this is the first work that covers the entire bias window. Our
findings include the characteristic slope of $G_{0}/2$
in the $I-V$ profile at the zero-bias limit when the two leads couple
symmetrically to the quantum dot, in accordance to the prediction
of Ref. {[}\onlinecite{del11}{]}. However, in the asymmetric case
($y\neq1$) we find a deviation from this slope, with $dI_{L}/dV>G_{0}/2$
and $dI_{R}/dV<G_{0}/2$ or the opposite, depending on the degree
of asymmetry. We also compare both $dI_{L}/dV$ and $dI_{R}/dV$ with
the conductance obtained via Eq. (\ref{Glinear}). They all agree
only for symmetric coupling ($y=1$). This indicates that a full nonequilibrium
quantum transport formulation is required to a better description
of the system. Our results were also compared to those
expected when a quantum dot is coupled to a RF zero-mode,
instead of a MBS. The two cases (RF and MBS) differ
appreciably in the entire bias-voltage range, not
only at the zero bias regime. Additionally, we observe the formation of a
plateau in the $I-V$ profile for intermediate bias voltages when the dot
is coupled to a MBS. This plateau is not seen in the RF case.
We also note that when the reservoirs temperature is large enough the two cases coincide, 
thus becoming indistinguishable via transport measurements if the dot is attached to a
RF level or to a MBS.

\section*{Acknowledgments}

This work was supported by the Brazilian agencies CNPq, CAPES, FAPEMIG, FAPESPA,
VALE/FAPESPA, ELETROBRAS/ELETRONORTE and PROPe/UNESP.


\begin{thebibliography}{20}
\bibitem{comment1}
In these equations we have used the definitions
\begin{equation}
  [i \frac{\partial}{\partial t}-\varepsilon_{k\a}] g_{k\a}(t,t') = \d(t-t'),
\end{equation}
and
\begin{equation}
  [i \frac{\partial}{\partial t}+\varepsilon_{k\a}] g'_{k\a}(t,t') = \d(t-t').
\end{equation}
\bibitem{comment2} In the next two expressions we write explicitly $\hbar$ or $h$.
\end{thebibliography}
\end{document}